# The strongest aftershock in seismic models of epidemic type

## G. Molchan[1]

1  Institute of Earthquake Prediction Theory and Mathematical Geophysics, Russian Academy of Science, 84/32 Profsoyuznaya st.,117997, Moscow, RF, Email molchan@mitp.ru

## E. Varini[2]

2  Institute for Applied Mathematics and Information Technologies "Enrico Magenes", National Research Council, via Corti 12, 20133 Milano, IT, Email elisa.varini@cnr.it

**Summary**.

We consider an epidemic-type aftershock model (ETAS(F)) for a large class of distributions F determining the number of direct aftershocks. This class includes Poisson, Geometric, Negative Binomial distributions and many other. Assuming an exponential form of the productivity and magnitude laws, we find a limiting distribution of the strongest aftershock magnitude $\mu_a$ when the initial cluster event $m_\bullet$ is large. The regime can be either subcritical or critical; the initial event can be dominant in size or not. In the subcritical regime, the mode of the limiting distribution is determined by the parameters of productivity and the magnitude laws; the shape of this distribution is not universal and is effectively determined by F. For example, the Geometric F- distribution generates the logistic law, and the Poisson distribution (studied earlier) generates the Gumbel type 1 law. The accuracy of these laws for moderate initial magnitudes is tested numerically. The limit distribution of the Båth's difference $m_\bullet - \mu_a$ is independent of the initial event size only if the regime is critical, and the ratio of exponents in the laws of magnitude and productivity is contained in the interval (1, 2). Previous studies of the $\mu_a$-distribution have dealt with the traditional Poisson F model and with arbitrary(not necessarily dominant) initial magnitude $m_\bullet$.

## 1 Introduction

**1.1 ETAS model**. The epidemic-type aftershock sequence model (ETAS) is a standard tool for assessing and analyzing seismicity [e.g., Ogata, 1988; Ogata and Zhuang, 2006; Nandan et al, 2021]. Traditionally, the model is defined by means of the conditional intensity of an event at a point of the phase space S= (time, location, magnitude), given the past of the seismic process up to the current time. The model in this form is convenient for modeling and forecasting, but is not flexible enough to describe direct aftershocks [Shebalin et al, 2020]. This shortcoming can be easily overcome by describing a cluster of seismic events in S as a Galton-Watson tree [Molchan et al, 2022]. In the future, the coordinates (time, place) are irrelevant. Therefore, we define the model in projection to the magnitude.

The initial event of magnitude $m$ generates a random number $v(m)$ of direct aftershocks $\{\mu_i\}$ according to the *off-spring law F*:

$$P(v(m) = k) = p_k(m). \qquad (1)$$

Each of these event is assigned an independent magnitude in accordance with the law $F_1$:

$$P(\mu \leq m) = \int_0^m f_1(x)dx =: F_1(m). \qquad (2)$$

Each event of the first generation independently generates new ones, following the described rule for its parent, etc. The process continues indefinitely. It stops with probability 1 if the *criticality index*

$$n = \int_0 \lambda(m) f_1(m) dm \leq 1, \qquad (3)$$

[Harris, 1963], where $\lambda(m) = Ev(m)$ is the *productivity* of the $m$-event. The values $n < 1$ and $n = 1$ correspond to the *subcritical* and *critical* regimes, respectively.

We get the standard ETAS model if F has a Poisson distribution (P), [e.g., Baró, 2020], that is

$$p_k(m) = \lambda^k(m) e^{-\lambda(m)} / k!. \qquad (4)$$

Shebalin et al [2020] found that the Geometric distribution, F=G:

$$p_k(m) = p^k(m)(1 - p(m)), \quad p(m) = \lambda(m)/(1 + \lambda(m)), \qquad (5)$$

agrees better with the empirical data.

Accepting this fact, the more general ETAS(F) models deserve attention. The choice of model F is not straightforward because of the difficulty of identifying direct aftershocks. The practical choice of model can be influenced by both the goals and the ease of modeling. A natural step in this situation is to expand the verification of the ability of models to reproduce various properties of real seismicity. For this purpose, we start from Båth's law, the analysis of which within the framework of ETAS(F) models is of independent interest.

**1.2 Båth's law**. The empirical Båth's law [Richter 1958; Båth, 1965], is related to the magnitude gap $\delta m = m_\bullet - \mu_a$ between the main event $m_\bullet$ and the strongest aftershock $\mu_a$. The law asserts that for shallow shocks the conditional mean value $E(\delta m | m_\bullet)$ does not depend of $m_\bullet$ and $E(\delta m|m_\bullet) \approx 1.2$. Later, Båth (1984) found that the gap increases slightly with increasing $m_\bullet$. Zaliapin and Ben-Zion[2013] showed the spatial variability cluster characteristics on the scale of tens of kilometers with respect to heat flow and other properties determining the effective viscosity of the region. This may explain the bimodality of the $\delta m$-distribution for circum-Pacific earthquakes (M7-8.4) with modes 1.2 and 1.8, [Tsapanos, 1990], and for Taiwan earthquakes (Ms > 5) with modes 0.3 and 0.9 [Chen, Wang, 2012]. An addition difficulty for Båth's law is the fact that substantial part (~20%) of global main shocks or main shocks in Japan turn out to be a part of doublets of earthquakes with difference not more than 0.4 units of magnitude and anomalously close in space-time (Kagan and Jackson, 1999; Grimm et al., 2023; Nandan et al.,2022)

The theoretical analysis of the $\mu_a$-distribution for clusters in the traditional ETAS(P) model was undertaken in [Saichev, Sornette ,2005; Zhuang, Ogata, 2006; Vere-Jones, 2008; Vere-Jones, Zhuang,2008; Luo & Zhuang,2016]. The analysis refers to the case when the initial event $m_\bullet$ is arbitrary (not necessarily dominant) and large, $m_\bullet >> 1$. In this case the limit $\mu_a$- distribution is doubly exponential, which is typical for the theory of extreme values.

**1.3 Random Thinning (RT) property**. Below we will significantly expand the theoretical analysis by considering a special class of distributions F having the so-called Random Thinning (RT) property [Renyi, 1956; Molchan et al, 2022]. Roughly speaking, the RT property assumes that the off-spring distribution after a random loss of a given proportion *p* of direct aftershocks changes only its average value $\lambda$ to $(1-p)\lambda$, but does not change its type, say *P* or *G* defined above.

The choice of F distributions with RT property is quite natural in the face of unavoidable errors in identifying direct aftershocks. The Generating function of such distribution has the following representation

$$\varphi_F(z|m) := Ez^{\nu(m)} = \phi_F(\lambda(m)(z-1)). \qquad (6)$$

The $\phi_F(w), w<0$ function uniquely determines the *type of the F-distribution*. Relationship (6) can be viewed as a formal definition of F-distributions with the RT property. The analytical properties of the $\phi_F(w)$ are described by Statement 3 in Appendix 1.

An important example of a distribution with the RT property is the Negative Binomial distribution F= $NB(\tau)$. Its type depends on the $\tau$-parameter since

$$\varphi_F(z|m) = (1-\lambda(m)(z-1)/\tau)^{-\tau}, \tau > 0. \qquad (7)$$

The case $\tau = 1$ corresponds to the Geometric distribution, while the limiting case $\tau = \infty$, for which $\varphi_F(z|m) = \exp(\lambda(m)(z-1))$, corresponds to the Poisson distribution. The class of distributions with RT property is quite wide. For example, the following operations on RT-types $\{\phi_i, i = 1,...,n\}$ :

$$\phi(w) = \prod_1^n \phi_i(p_i w) \quad \text{and} \quad \phi(w) = \sum_1^n \phi_i(w) p_i \;, \text{ where } \quad \sum_1^n p_i = 1, \; p_i > 0 \qquad (8)$$

allow to obtain new types of distributions with RT property [Molchan et al, 2022].

**1.4 The problem.** Båth's law can be interpreted in different ways, depending on the cluster type In one type of clusters, the initial $m_\bullet$ event may be the strongest (dominant), in another type, the $m_\bullet$ event may be arbitrary. Therefore, we will consider clusters of both types: $DM(m_\bullet)$ with Dominant initial Magnitude and $AM(m_\bullet)$ with Arbitrary initial Magnitude. Our task is to find in these cases the distribution of the maximum aftershock magnitude under the condition $m_\bullet >> 1$. The case with the dominant initial magnitude refers to the original interpretation of Båth's law.

The main result and its discussion can be found in Sections 2.2 and 2.3. The actual accuracy of the limit laws and the difficulty of choosing a model F are briefly discussed in Sections 2.4, 2.5. All auxiliary mathematical statements and proofs are placed in the Appendix. Below we retain the above designations for magnitudes: the random quantities $\mu$, $\mu_a$ (strongest aftershock), $\delta m$ (Båth's gap) start with Greek letters, while the deterministic ones $m$, $m_\bullet$ (initial magnitude), $M$ start with Latin letters.

## 2. Theoretical results

The probabilistic structure of $DM(m_\bullet)$ cluster in the ETAS(F) model is given by Statement 4 in Appendix 2. An ETAS(F) cluster of type $DM(m_\bullet)$ can be viewed as an $AM(m_\bullet)$ cluster in the ETAS($\hat{F}$) model with a suitable off-spring distribution $\hat{F}$.

**2.1 The main equation.**

We will consider clusters in which the number of all events, including the initial one, is at least two, i.e. the initial event $m_\bullet$ and the strongest aftershock $\mu_a$ are different as events.

**Statement 1.** a) Consider AM-cluster in the ETAS(F) model with the generating function $\varphi_F(z|m)$ of the distribution F. Let $M := M(\pi)$ be the quantile of the level $\pi \in (0,1)$ for the distribution of $\mu_a$, that is $P(\mu_a < M) = \pi$. If $\psi := \psi(\pi)$ is defined by the following equation

$$\pi = \frac{\varphi_F(\psi|m_\bullet) - \varphi_F(0|m_\bullet)}{1 - \varphi_F(0|m_\bullet)}, \qquad (9)$$

then M is given by the equation :

$$\psi(\pi) = \int_0^M f_1(m)\varphi_F(\psi(\pi)m)dm . \tag{10}$$

**Remark.** In the case $F = NB(\tau), 0 < \tau \leq \infty$ function (9) is explicitly reversible as follows

$$\psi(\pi) = 1 - \{[\pi + (1-\pi) \cdot p_0(m_\bullet)]^{-1/\tau} - 1\}\tau / \lambda(m_\bullet), \tag{11}$$

where $p_0(m_\bullet) = (1 + \lambda(m_\bullet)/\tau)^{-\tau}$.

The $DM(m_\bullet)$ cluster in the ETAS($NB(\tau)$)-model can be considered as an AM-cluster in the same type of model but with new $(f_1(m), \lambda(m))$ characteristics:

$$\hat{f}_1(m) = f_1(m) / F_1(m_\bullet), 0 \leq m \leq m_\bullet, \tag{12}$$

$$\hat{\lambda}(m) = \lambda(m) F_1(m_\bullet) / (1 + \lambda(m)\tau^{-1}(1 - F_1(m_\bullet))). \tag{13}$$

## 2.2 The $\mu_a$-distribution for $m_\bullet \gg 1$.

Under $m_\bullet \gg 1$ condition, equation (10) allows us to obtain the limiting magnitude distribution of the strongest aftershock for both types of clusters: with a dominant (DM) or arbitrary (AM) initial magnitude. It is described by the following

**Satement 2.** Consider ETAS(F) model with a distribution $F$ having the RT property, that is

$$Ez^{\nu(m)} = \phi_F(\lambda(m)(z-1)), \lambda(m) = E\nu(m) . \tag{14}$$

Suppose that

a) $P(\nu(m) = 0) \to 0, m \to \infty$,     b) $E\nu^2(m) < \infty$ , $\tag{15}$

c) $\lambda(m) = \lambda_0 e^{\alpha \cdot m}, m \geq 0$,     d) $f_1(m) = \beta e^{-\beta m}/(1 - e^{-\beta M_1}), 0 \leq m \leq M_1 = Km_\bullet$, $\tag{16}$

where $\alpha \leq \beta$, $1 \leq K < \infty$; if $\alpha < \beta$, $K \leq \infty$.

Let $\mu_a$ be a maximal magnitude in the AM/DM cluster and $n$ is the criticality index.

**A.** Under $m_\bullet \gg 1$ condition, the following regression relationships are valid:

$$\beta\mu_a = \alpha m_\bullet + \ln(\lambda_0/(1-n)) + \zeta \quad (\alpha < \beta, n < 1); \tag{17}$$

$$\beta\mu_a = \alpha m_\bullet + \ln(\lambda_0/(1-n/K)) + \zeta \quad (\alpha = \beta, n \leq 1) \tag{18}$$

(since $n$ is fixed, $\lambda_0$ in (18) is a function of $m_\bullet$: $\lambda_0 = n(\beta K)^{-1}/m_\bullet$);

$$\beta\mu_a = 2\alpha m_\bullet - \ln m_\bullet \cdot 1_{\beta=2\alpha} + \ln A + 2\zeta \quad (2\alpha \leq \beta, n = 1), \tag{19}$$

where $A = 2(\beta\phi_F''(0))^{-1}(\beta - 2\alpha + 1_{\beta=2\alpha})$.

The random component in (17-19) has the limit distribution

$$P(\zeta < x) = \phi_F(-e^{-x}), |x| < \infty \tag{20}$$

regardless of the cluster type DM or AM.

For *x*>0, relation (20) has the following meaning:

$$P(\zeta < \beta x) = \phi_F(F_1(x) - 1)$$
$$= P\{\text{ all direct aftershocks of the m event with } \lambda(m) = 1 \text{ have the magnitude } < x\}. \quad (21)$$

**B.** In the case *n*=1 and $2\alpha > \beta > \alpha$, the limit $\mu_a$-distribution depends on the function $V(x), x > 0$, defined by the following equation

$$V + (\beta/\alpha - 1)V^{\beta/\alpha} \int_0^V u^{-\beta/\alpha - 1}[\phi_F(-u) - 1 + u]du = x. \quad (22)$$

**B1**. For AM- cluster,

$$\alpha\mu_a = \alpha m_\bullet + \ln V(\lambda_0) + \zeta, \quad (23)$$

where the random component $\zeta$ has the limit distribution (20).

Since $V(x) \approx x$ for small $..1 - \alpha/\beta = \lambda_0$, in this case we will have

$$P(\mu_a - m_\bullet < x) \approx \begin{cases} 1 - \lambda_0 e^{-\alpha x}, & x > 0 \\ \phi_F(-\lambda(|x|)), & x < 0 \end{cases} \quad (24)$$

**B2**. For DM cluster, the limit $\mu_a$ distribution is

$$P(\mu_a - m_\bullet < x) = \phi_F(-V(\lambda_0(1 - e^{\beta x}))e^{-\alpha x}), \quad x < 0 \quad (25)$$

For small $\lambda_0 \approx 1 - \alpha/\beta$

$$P(\mu_a - m_\bullet < -x) \approx \phi_F(-\lambda(x)F_1(x)), \quad x \geq 0.$$

**2.3 Comments.**

1) *Critical parameters*.

Parameters $(\alpha/\beta, n) \in (0,1] \times (0,1] = U$ are essential to describe the limit $\mu_a$ distribution. The critical points $\alpha/\beta \in \{1/2; 1\}$ and $n = 1$ divide U into 5 fragments:

$$U_1 = (0,1) \times (0,1); \quad U_2 = 1 \times (0,1]; \quad U_3 = (0,1/2) \times 1; \quad U_4 = 1/2 \times 1; \quad U_5 = (1/2,1) \times 1$$

in which the normalization of $\mu_a$ and mode of the limit $\mu_a$-distribution are continuous across the parameters in the problem. The critical values of the $\alpha/\beta$ parameter are due to the properties of the $v(\mu)$ statistic when the initial event $\mu$ is random and has a non-truncated exponential distribution: as $\alpha/\beta$ increases, the first and second moments of this statistic become infinite, starting at $\alpha/\beta = 1$ and $\alpha/\beta = 1/2$, respectively.

The U1, U3, U5 regimes have been previously studied in [Saichev, Sornette, 2005; Zhuang, Ogata, 2006; Vere-Jones, 2008; Vere-Jones, Zhuang, 2008; Luo & Zhuang, 2016]. These works are related to AM clusters within the traditional ETAS(*P*) model.

2) *Non-random of $\mu_a$-regression component*.

In the subcritical regime, $n < 1$, the non-random component in (17) can be represented as follows

$$\overline{m}_a = \beta^{-1} \ln[\lambda_0 e^{\alpha m_\bullet}/(1-n)] \approx \beta^{-1} \ln \overline{N}(m_\bullet), \qquad (26)$$

where $\overline{N}(m_\bullet)$ is the average size of the cluster with the initial $m_\bullet$-event (not necessarily dominant). As (Helmstetter, Sornette, 2003) noted,

$$\beta^{-1} \ln \overline{N}(m_\bullet) = E \max(\mu_1,...,\mu_{[\overline{N}(m_\bullet)]})(1+o(1)), m_\bullet \gg 1, \qquad (27)$$

where $\{\mu_i\}$ are independent variables with the $m$-distribution $f_1(m) = \beta e^{-\beta m}$. That is the average number of cluster events well predicts the mode in the $m_a$ distribution. However, in the critical regime ($n=1$) or $\alpha = \beta$ this property disappears.

The universality (relative to F) of the $\overline{m}_a$ component in sub-critical regime is due to the fact that the average cluster size with initial magnitude $m_\bullet$ is determined by the laws of magnitude and productivity common to all F-models considered.

3) *The random component of the $\mu_a$-regression.*

The distribution of the random component of the $\mu_a$ regression is not universal and is determined by the off-spring F law. In the case $F=NB(\tau)$, the limit distribution (20) is known as the generalized Logistic distribution [Gupta, Kundu, 2010]:

$$P(\zeta < x) = (1 + e^{-x}/\tau)^{-\tau}, |x| < \infty \qquad (28)$$

with the mode at $x=0$. The standard Logistic distribution corresponds to the case $\tau = 1$, i.e. to the Geometric distribution, F=G. In the Poisson case ($\tau = \infty$) we have the Gumbel type-1 distribution, $P(\zeta < x) = \exp(-e^{-x})$ (see e.g. Johnson et al, 1995). Its truncated analogue was used in (Grimm et al, 2023), although the authors proceeded from the hypothesis that the off-spring distribution is $F=NB(\tau = 2)$.

The density of $\phi_F(-e^{-\beta x})$-distribution has a maximum at point 0 if $\phi_F''(-1) = \phi_F'(-1)$. This is true for the case related to the Negative-Binomial distribution of the direct aftershocks. Therefore, the non-random component of $\mu_a$ in the regression relations (17-19,23) for ETAS($F=NB(\tau)$) models can be interpreted as the mode of the $\mu_a$ distribution.

Despite the diversity of the random components, in the subcritical regime, the distributions of $\zeta/\beta$ have a *universal exponential behavior* at large values:

$$P(\zeta/\beta > x) = 1 - \phi_F(-e^{-\beta x}) \approx e^{-\beta x}, x \to \infty. \qquad (29)$$

However, the left tail of this distribution loses versatility. In the model with the Negative Binomial distribution, $F=NB(\tau)$, the left $\zeta$-tail can be both stronger ($\tau > 1$) and slower ($\tau < 1$) than the standard one, (29). In the model with a geometric distribution $(\tau = 1)$, the $\zeta$ - distribution is symmetric. Therefore, for $x<0$, the scatter of the noise component for F=G and F=P is visually different, because it is roughly described by tails of the form

$$\exp(-|x|) \quad \text{vs} \quad \exp(-e^{|x|}) \quad , x<0 \; .$$

This fact can be useful to test F.

4) *DM and AM clusters*.

The probabilistic structure of a cluster with a dominant initial event is also described by the ETAS model (see Appendix 2). Under condition $m_\bullet >> 1$ the $(f_1(m), \lambda(m))$ characteristics of DM and AM clusters become close. Therefore, it is quite expected that the limit distributions of the maximum aftershock in both cases will be the same. These intuitive considerations proved themselves in the subcritical regime and even in the critical one, if $\alpha = \beta$ or $2\alpha \leq \beta$. In the critical regime, $n=1$, the $\mu_a$ - distributions for the two types of clusters are radically different from each other if $\alpha < \beta < 2\alpha$. Some simplification is possible for small $\lambda_0 = 1 - \alpha/\beta$. In the case F=NB$(\tau)$, $0 < \tau \leq \infty$, the $\mu_a$ distribution looks as follows:

$$P(\mu_a - m_\bullet < x) \approx \begin{cases} 1 - \lambda_0 e^{-\alpha x} & x > 0 \\ (1 + \lambda_0 e^{-\alpha x}/\tau)^{-\tau}, & x < 0 \end{cases} \quad \text{(AM- cluster)} \qquad (30)$$

$$P(\mu_a - m_\bullet < x) \approx (1 + \lambda_0 e^{-\alpha x}(1 - e^{\beta x})/\tau)^{-\tau}, \; x < 0 \quad \text{(DM-cluster)} \qquad (31)$$

5) *The Båth's law*

The mode of the Båth's gap $\delta m = m_\bullet - \mu_a$ grows linearly $\hat{\delta m} = (1 - \alpha/\beta)m_\bullet + C$ with $m_\bullet$ if $\alpha < \beta, n < 1$ and logarithmically, $\hat{\delta m} = \beta^{-1} \ln m_\bullet + C_1$ if $(\alpha = \beta, n \leq 1)$ or $(2\alpha = \beta, n = 1)$. Felzer et al.(2004) considered the $\alpha = \beta$ condition as necessary for the independence of the Båth's gap from the magnitude of the main event. If we do not neglect the logarithmic dependence of $\hat{\delta m}$ on $m_\bullet$, then condition $\alpha = \beta$ is not sufficient. Moreover, this condition is not necessary for all our models. The limit distribution of the Båth's difference $m_\bullet - \mu_a$ is independent of the initial event size only if the regime is critical, $n=1$, and $\alpha < \beta < 2\alpha$. This fact is unexpected. The necessity of condition n=1 was previously noted in [Saichev, Sornette, 2005].

## 2.4 Numerical testing

The theoretical analysis of the $\mu_a$-distribution is obtained for $m_\bullet >> 1$ and therefore requires verification for moderate values of $m_\bullet$. To test the $\mu_a$-distribution, we considered ETAS(F) models with Poisson F=P and geometric F=G distributions, as well as clusters with arbitrary (AM) and dominant (DM) initial magnitude $m_\bullet$ under conditions $(\alpha < \beta, n < 1)$. The method for calculating the exact $\mu_a$- distributions is described in Statement 1.

*Numerical parameters.* The initial magnitude $m_\bullet$ varied in the range 2-6, assuming that the lower magnitude threshold is taken as 0. The parameters of the productivity and magnitude laws are similar to those used by Helmstetter and Sornette [2003] for California:

$$\alpha = 1.8, \quad \beta = 2.08 \text{ (b-value=1)}, \, n = 0.7.$$

Figure 1 shows the $\delta m = m_\bullet - \mu_a$-distribution densities for two F- models (P/G) and two cluster types (AM/DM). In the limiting case, the $\delta m$-distribution is symmetric for the case F=G and asymmetric for F=P, and the left tail of the distribution is exponential. These features can also be clearly seen in the prelimit distributions, especially in the case of AM clusters.

The limit $\mu_a$-distribution, $\pi_\infty(M)$, is determined by the noise limit distribution in the regression (17):

$$F_\varsigma(x) = P(\varsigma < x) = \phi_F(-e^{-x}), |x| < \infty.$$

Using the inverse function $\bar{F}_\varsigma(x)$ we get the linear relation

$$-\bar{F}_\varsigma(\pi_\infty(M)) = -\beta M + \alpha m_\bullet + \ln(\lambda_0/(1-n)). \tag{32}$$

Let's replace $\pi_\infty(M)$ with the exact distribution $\pi(M)$ of $\mu_a$-statistics. The curves W(M)= $-\bar{F}_\varsigma(\pi(M))$ for all options under consideration are shown in Figure 2. In doing so, we used the $\pi_\infty(M)$ distribution variants (see (A20,A34)), which takes into account that the cluster size is at least 2. In this case the support of $d\pi_\infty(M)$ measure is in the interval M> $M_-$.

In Figure 2, the differences between models F= G and F= P disappear in the region of central values of distributions and the expected linearity of regressions (32) are well reproduced. The regression parameters themselves are close for the AM and DM clusters, but may differ from the limit values (see Table 1). All this suggests that the result about the noise component in the limit $\mu_a$-distribution is quite acceptable for the real application.

(a)                                                            (b)

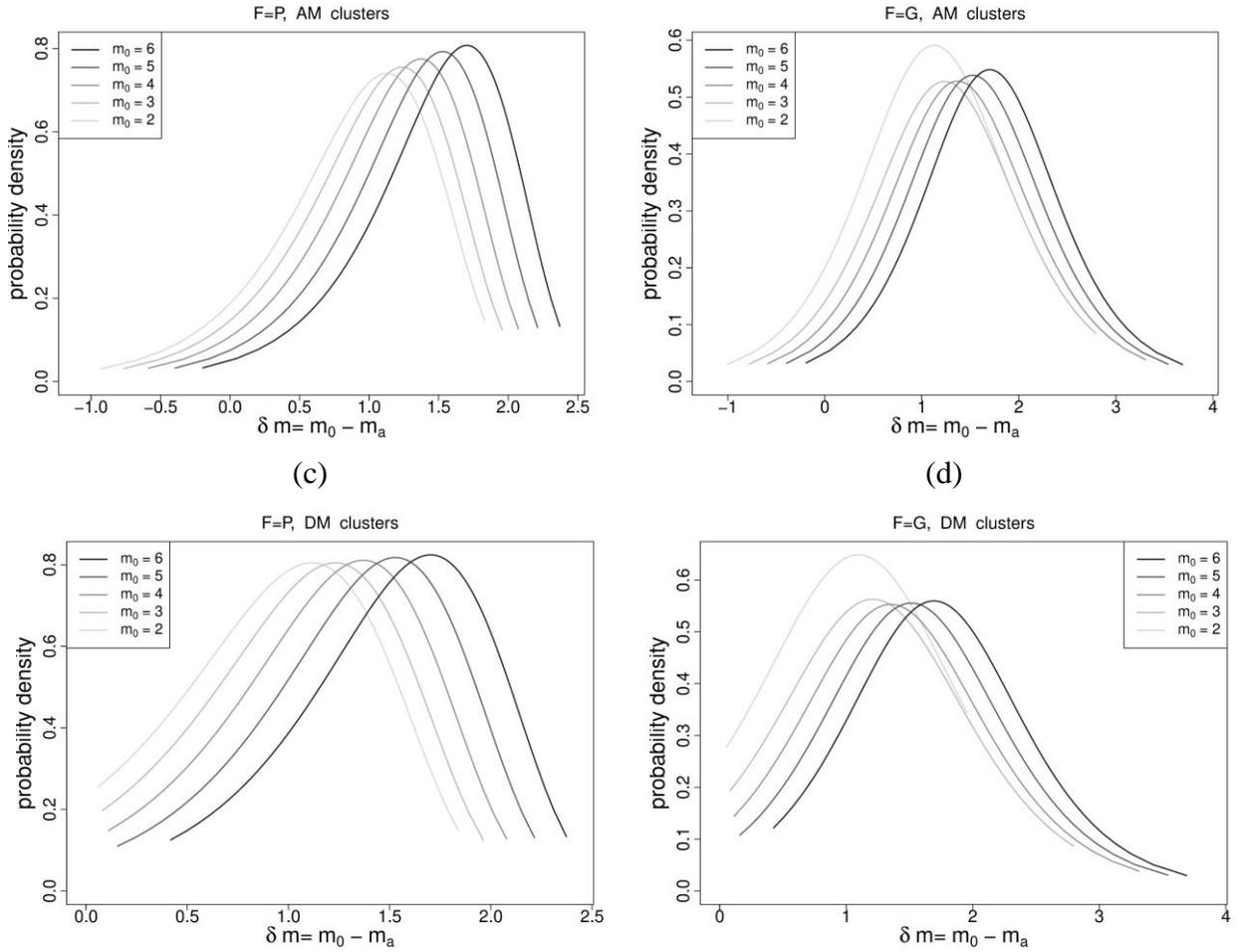

**Figure 1.** Distribution density of the Båth's magnitude difference $\delta m = m_\bullet - \mu_a$, depending of the initial magnitude $m_\bullet = 2-6$, AM/DM cluster type, and off-spring distribution $F$: Poisson($P$) and Geometric ($G$).

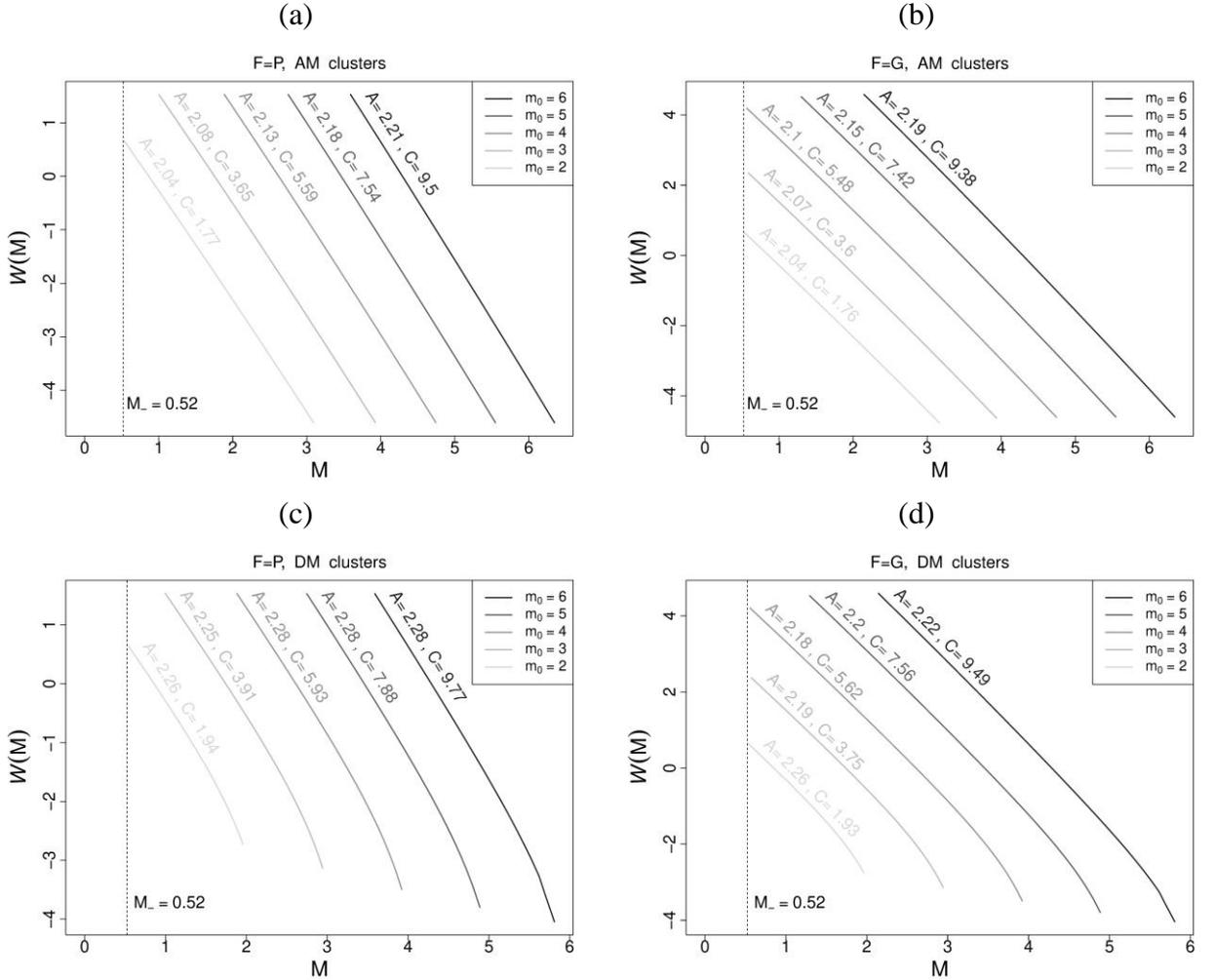

**Figure 2.** Noise component test, depending of the initial magnitude $m_0 = 2-6$, AM/DM cluster type, and off-spring distribution *F*: Poisson(*P*) and Geometric (*G*). In the limit case, $W(M) = -A_\infty M + C_\infty = -\beta M + \alpha m_\bullet + \ln \lambda/(1-n)$. The prelimit (*A,C*) coefficients are shown in the figures.

|  | AM clusters | | DM clusters | | |
|---|---|---|---|---|---|
|  | F = P | F = G | F = P | F = G |  |
| $m_0$ | $A$ | $A$ | $A$ | $A$ | $A_\infty$ |
| 2 | 2.04 | 2.04 | 2.11 | 2.16 | 2.30 |
| 3 | 2.08 | 2.07 | 2.15 | 2.14 | 2.30 |
| 4 | 2.13 | 2.10 | 2.21 | 2.16 | 2.30 |
| 5 | 2.18 | 2.15 | 2.26 | 2.20 | 2.30 |
| 6 | 2.21 | 2.19 | 2.28 | 2.25 | 2.30 |
|  | AM clusters | | DM clusters | | |
|  | F = P | F = G | F = P | F = G |  |
| $m_0$ | $C$ | $C$ | $C$ | $C$ | $C_\infty$ |
| 2 | 1.77 | 1.76 | 1.81 | 1.83 | 2.91 |
| 3 | 3.65 | 3.60 | 3.73 | 3.68 | 4.71 |
| 4 | 5.59 | 5.48 | 5.76 | 5.58 | 6.51 |
| 5 | 7.54 | 7.42 | 7.79 | 7.56 | 8.31 |
| 6 | 9.50 | 9.38 | 9.77 | 9.59 | 10.11 |

**Table 1.** ($A,C$) coefficients from Fig. 2 and their limit values $A_\infty = \beta$ and $C_\infty = \alpha m_\bullet + \ln(\lambda/(1-n))$,

### 2.5 Selecting ETAS(F) model.

The universal nature of a number of properties of the limiting distribution of the strongest aftershock makes it difficult to choose the most preferable ETAS(F) model in relation to Båth's law. Above we noted a practically useful but not strong enough feature to distinguish F=P model from F=G: this is the symmetry and asymmetry of the $\mu_a$ distribution for F=G and F=P, respectively. Therefore, let us turn to an important characteristic, <d>, of an epidemic-type seismic model introduced by Zaliapin and Ben-Zion [2013], which we will call average *cluster depth* (instead of average *leaf depth*).

To give a definition, let us consider a chain of events connected by causal relationships. Such a chain begins with the cluster initial event, each subsequent event in the chain is a direct aftershock of the previous one, and the last event has no aftershocks. The number of the chain events determines its 'depth', *d*, and their empirical average value over all chains in the cluster is the statistics <d>. Next, we will assume that the depth refers to the conditional situation when the cluster contains at least two events.

Obviously, <d> is closely related to the 0-probability $p_0(m) = P(\nu(m) = 0)$ in the off-spring distribution *F* since $p_0(m)$ controls the chain condition. For ETAS(*F*) models (see Appendix 5)

$$E<d> = Ed = [\int_0^\infty p_0(m) f_1(m) dm]^{-1} \leq 1/(1-n), \tag{33}$$

where $n$ is the criticality index. The upper bound in (33) is sharp estimate of $Ed$ for the whole class of ETAS models; in particular, $Ed \leq 5$ if n=0.8.

For ETAS($F = NB(\tau)$) model with arbitrary ($f_1(m), \lambda(m)$) characteristics, the function $\tau \to Ed$ increases; at that

$$Ed < e^n, \quad 0 < \tau \leq \infty. \tag{34}$$

In this case $Ed \leq 2.2$ if n=0.8.

Hence, by replacing the Poisson distribution $F=P$ in ETAS model with any $F=NB(\tau)$ we reduce the average cluster depth.

For all the difficulties of empirically estimating $<d>$, analysis of the California data [Zaliapin,Ben-Zion,2013] showed that the range of $<d>$ is much wider in real catalogue than in the synthetic one, based on ETAS($P$) model; namely, (1, 25) vs. (1, 5). The authors concluded that ETAS models do not fully describe the mechanism of cluster formation. At the same time, the results of [Zaliapin, Ben-Zion,2013] do not exclude the admissibility of the ETAS($P$) model to describe seismicity in regions with low heat flow.

Thus, the statistic $<d>$ testifies in favor of choosing the traditional Poisson model from the family with $F= NB(\tau)$) while the data, related of the off-spring distribution $F$, testifies to the opposite. This means that the $p_0(m)$ probability requires additional analysis.

What happens if we change 0-probability in the $F = \{p_k(m)\}$ model as follows:

$$\tilde{F}: \tilde{p}_0(m) = \varepsilon(m), \quad \tilde{p}_k(m) = p_k(m)(1-\varepsilon(m))/(1-p_0(m)), k \geq 1, \tag{35}$$

This operation will preserve the shape of $F$ for k>0 and increase the average cluster depth provided that $\varepsilon(m) < p_0(m)$ and the new criticality index does not exceed 1:

$$\tilde{n} = \int_0^\infty \lambda(m) f_1(m)(1-\varepsilon(m))/(1-p_0(m)) dm \leq 1. \tag{36}$$

In the case $F = NB(\tau)$ we will have $\tilde{n} < 1$ and $E\tilde{d} = Ed/n$ if $\varepsilon(m) = np_0(m)$, $\tau > 1$, and n<1. In the case $\tau = 1$, we have $\tilde{n} = 1$.

The most of the conclusions about the limit distribution of $\mu_a$ in the ETAS($F$) model remain valid for the ETAS($\tilde{F}$) model as well. In particular, this applies to the relations (17-19) for AM clusters, in which the index $n$ should be replaced by $\tilde{n}$ and another constant $A$ should be used if $\beta \neq 2\alpha$. The proof literally repeats the arguments for the case $F$.

## Conclusion

Within the ETAS(*F*) model with regular the magnitude and productivity characteristics, we obtained limit distributions for the strongest aftershock $\mu_a$ of the initial event $m_\bullet \gg 1$. The problem was solved depending on the type of criticality of the regime, the dominance of the initial event size, and the off-spring distribution *F*.

In the subcritical regime, the mode of the limiting distribution is determined by the parameters of productivity and the magnitude law; the shape of this distribution is not universal and is effectively determined by *F*.

The limit distribution of the Båth's gap $\delta m = m_\bullet - \mu_a$ is independent of the initial event only if the regime is critical, and the ratio of exponents in the laws of productivity and magnitude, $\alpha/\beta$, is contained in the interval (1/2, 1). We have identified a number of universal properties of the limit $\delta m$- distribution; in particular, it's the mode and the exponential tail of $\delta m$ distribution to the left of the $E\delta m$ value. For the practical choice of model *F*, the contrast of the scatter of the data in the area of large and small values of $\delta m$ is important. For example, in the Poisson model, the probability of large $\delta m$ values sharply decreases and the $\delta m$-distribution is asymmetric, whereas for the geometric model we have symmetry of the $\delta m$-distribution and the exponential tails.

Numerical testing of the noise component of the $\mu_a$ distribution has shown that it is possible to transfer the conclusions for large initial events $m_\bullet$ to moderate $m_\bullet \geq 3$.

Staying within the ETAS(*F*) model, the choice between Poisson and Geometric *F*- distributions remains controversial. This problem requires an analysis of the zero probability $p_0(m)$ in the distribution *F*.

**Acknowledgements.** The authors dedicate this article to the honorable memory of Ilya Zaliapin (1973–2023).

We are grateful to the reviewer Prof. D. Sornette for appreciating our work and to an anonymous reviewer for constructive comments.

## Appendix 1.

**Statement 3.** Let $v_\lambda, \lambda > 0$ be a family of random variables with the distribution: $P(v_\lambda = n) = p_n(\lambda), n \geq 0$ such that

$$p_0(\lambda) \to 0, \lambda \to \infty \text{ and } Ev_\lambda^2 < \infty .$$

Assume that for any $\lambda > 0$

$$\varphi(z|\lambda) := Ez^{v_\lambda} = \phi(\lambda(z-1)), |z| < 1, \qquad \lambda = Ev_\lambda$$

Then

a) $\phi(w), w < 0$ function is analytic, increasing from $\phi(-\infty) = 0$ to $\phi(0) = 1$;

b) $0 \leq \phi'(w) < \phi'(0) = 1$ ; $0 \leq \phi''(w) < \phi''(0) < \infty$ ;

c) $0 \leq \phi(w) - 1 - w < c(w^2 \wedge |w|)$.

**Proof**. By definition, $\varphi(z|\lambda)$ is analytical in the circle $|z| \leq 1$; it entails analyticity of $\phi_F(w)$ in the interval $-2\lambda < w < 0$. Since $\lambda > 0$ is arbitrary, $\phi_F(w)$ is analytical on the entire semi-axis $w < 0$.

Since $p_n(\lambda) \geq 0, n \geq 0$, $Ez^{v_\lambda} = \phi(\lambda(z-1))$ is an increasing function of $z \in (0,1)$. Hence $\phi(w)$ has the same property for $-2\lambda < w < 0$, and therefore for any $w < 0$.

Since $\phi(-\lambda) = \varphi(0|\lambda) = p_0(\lambda) \to 0, \lambda \to \infty$, we have $\phi(-\infty) = 0$. In addition, $\phi(0) = \varphi(1) = 1$.

Similarly, $\varphi^{(k)}(z|\lambda) = E\prod_{i=0}^{k-1}(v(\lambda) - i)z^{v(\lambda)-k}$ is an increasing function of $z \in (0,1)$, since the product under the expectation sign is non-negative at integer values of $v(\lambda)$. As above, we can conclude, that $\phi^{(k)}(w) \geq 0$. The existence of derivatives $\phi^{(k)}(0), k = 1, 2$ ensures their continuity at point 0. We have $\lambda = \varphi'(1|\lambda) = \lambda\phi'(0) \geq 0$, hence $\phi'(0) = 1$.

Let us estimate the function $\psi(w) = \phi(w) - 1 - w$. This function is convex, since $\psi''(w) = \phi''(w) \geq 0$. Hence $\psi'(w)$ increases up to $\psi'(0) = \phi'(0) - 1 = 0$, that is $\psi'(w) \leq 0$. But then $\psi(w) \geq \psi(0) = 0$.

Next,
$$\psi(w) = \phi(w) - 1 - w \leq 1 - 1 + |w| = |w|,$$
and
$$\psi(w) = \phi(w) - 1 - w = \phi''(\theta w)w^2/2 \leq \phi''(0)w^2/2 = Cw^2,$$
where $\theta = \theta(w) \leq 1$. The lemma is proven.

## Appendix 2.

**Statement 4.** Consider ETAS(F) model with magnitude and productivity characteristics $(f_1(m), \lambda(m))$. Let $\varphi_F(z|m)$ be the generating function for direct aftershocks of an *m*-event. Then $DM(m_\bullet)$ cluster in the ETAS(F) model can be viewed as $AM(m_\bullet)$ cluster in the ETAS($\hat{F}$) model such that the magnitude law has the form

$$\hat{f}_1(m) = f_1(m)/F_1(m_\bullet), 0 \leq m \leq m_\bullet \tag{A1}$$

and the off-spring law $\hat{F}$ is contiguous with F:

$$\hat{F}: \hat{p}_k(m) = p_k(m)F_1^k(m_\bullet)/\varphi_F(F_1(m_\bullet)|m). \tag{A2}$$

The generating function looks as follows

$$\varphi_{\hat{F}}(z|m) = \varphi_F(zF(m_\bullet)|m) / \varphi_F(F(m_\bullet)|m).$$

In the case $F = NB(\tau)$, any $DM(m_\bullet)$ cluster in the ETAS(F) model can be viewed as an $AM(m_\bullet)$ cluster of the same model, but with the new magnitude law (A1) and the productivity law

$$\hat{\lambda}(m) = \lambda(m)F_1(m_\bullet)/(1 + \lambda(m)\tau^{-1}(1 - F_1(m_\bullet))), \quad \tau > 0.$$

**Proof.** Consider a DM($m_\bullet, F$) cluster in the ETAS($F$) model. The structure of branching and independence in this cluster is the same as in the original model, except for the limitation on the magnitude: any event $m \leq m_\bullet$ generates new events with magnitude $< m_\bullet$. That is, direct aftershocks $\{\mu_i\}$ of any initial $m$-event realized under the condition $\Omega_0(m) = \{\mu_i < m_\bullet, i = 1,.., \nu(m)\}$. The probability of $\Omega_0(m)$ is

$$P(\Omega_0(m)) = \sum_0^\infty p_k(m)F_1^k(m_\bullet) = \varphi_F(F_1(m_\bullet)|m).$$

Therefore the conditional probability density of the direct aftershocks is

$$P\{\mu_i = m_i, i = 1,.., n, \nu(m) = n | \Omega_0(m)\} = \hat{p}_k(m)\hat{f}_1(m_1)...\hat{f}_1(m_k),$$

where $\hat{f}_1(m), \hat{p}_k(m)$ are given by (A1) and (A2). Hence

$$\varphi_{\hat{F}}(z|m) = \sum_0^\infty \hat{p}_k(m)z^k = \sum_0^\infty p_k(m)[zF(m_\bullet)]^k / P(\Omega_0(m))$$

$$= \varphi_F(zF_1(m_\bullet)|m) / \varphi_F(F_1(m_\bullet)|m).$$

In the case $F = NB(\tau)$ this relation looks as follows

$$(1 - \lambda(m)(zF_1(m_\bullet) - 1)/\tau)^{-\tau} / (1 + \lambda(m)\overline{F}_1(m_\bullet)/\tau)^{-\tau} = (1 - \hat{\lambda}(m)(z - 1)/\tau)^{-\tau}$$

where

$$\hat{\lambda}(m) = \lambda(m)F_1(m_\bullet)/(1 + \lambda(m)\tau^{-1}(1 - F_1(m_\bullet))).$$

Hence, the F law $\{\hat{p}_k(m)\}$ is again the NB($\tau$) distribution with the new average value $\hat{\lambda}(m)$.

## Appendix 3. Proof of Statement 1

Let $ETAS(F)$ be the cluster model with characteristics $(\lambda(m), f_1(m))$ and generating function $\varphi_F(z|m) := Ez^{\nu(m)}$ Consider the event

$A_m(M)$: all (direct and indirect) off-springs of initial $m$ event have a magnitude $<M$.

and its probability

$$z(M|m) := P(A_m(M)) \tag{A3}$$

Specifically, $z(M|m_\bullet) = P(\mu_a < M|m_\bullet)$, where $m_\bullet$ is the initial cluster event.

Let

$$z(M) = \int_0^M z(M|m) f_1(m) dm. \tag{A4}$$

We now find the equations for $z(M)$ and $z(M|m_\bullet)$.

Let $\{\mu_1, ..., \mu_n, \nu = n\}$ be the direct off-springs of a $m$-event, , then their probability density is

$$P\{\mu_1 = m_1, ... \mu_n = m_n, \nu = n|m\} = p_n(m) f_1(m_1)...f_1(m_n),$$

Since

$$A_m(M) = \bigcup_{n \geq 0} \{A_{\mu_1}(M), ..., A_{\mu_n}(M), \nu(m) = n\}$$

and $A_{\mu_1}(M), ..., A_{\mu_n}(M)$ are independent, we have

$$P\{A_{\mu_1}(M), ..., A_{\mu_n}(M), \nu = n|m\} = p_n(m) \prod_1^n \int P(A_{\mu_i}(M)) f_1(\mu_i) d\mu_i.$$

Finally

$$z(M|m) = \sum_{n \geq 0} p_n(m) z^n(M) = \varphi_F(z(M)|m)). \tag{A5}$$

By definition, the number of events in a cluster is greater than 1. Therefore, considering the $A_{m_\bullet}(M)$ event for the initial event $m_\bullet$, we must take into account that $m_\bullet$ has a positive number of events of first generation, that is $\nu(m_\bullet) > 0$. Therefore (A5) specially for $m = m_\bullet$ should be changed as follows

$$z(M|m_\bullet) = \sum_{n \geq 1} p_n(m_\bullet) z^n(M) / (1 - p_0(m_\bullet))$$
$$= [\varphi_F(z(M)|m_\bullet) - p_0(m_\bullet)] / (1 - p_0(m_\bullet)). \tag{A6}$$

Integrating (A5) by $m$ with weight $f_1(m)$, we get the second equation:

$$z(M) = \int_0^M f_1(m) \varphi_F(z(M)|m) dm. \tag{A7}$$

Let $M := M(\pi)$ be the $\pi$-quantile of the $\mu_a$-distribution., then $z(M|m_\bullet) = \pi$, and (A6) is converted into an equation with respect to the function $\psi(\pi) := z(M(\pi))$:

$$\pi = \frac{\varphi_F(\psi|m_\bullet) - \varphi_F(0|m_\bullet)}{1 - \varphi_F(0|m_\bullet)}. \tag{A8}$$

Substituting $z(M) = \psi(\pi)$ into (A7), we get a closed equation for $M := M(\pi)$

$$\psi(\pi) = \int_0^M f_1(m) \varphi_F(\psi(\pi)|m) dm$$

# Appendix 4.
## Proof of Statement 2.
### 4.1 AM($m_\bullet$, $F$) cluster.

Since $m_\bullet \gg 1$, we can simplify some elements of the ETAS(F) model:

- the criticality index

$$n = \int_0^{M_1} f_1(m)\lambda(m)dm = \lambda_0 \beta(1-e^{-(\beta-\alpha)\cdot M_1})/(\beta-\alpha) \approx \begin{cases} \lambda_0 \beta/(\beta-\alpha), & \beta > \alpha \\ \lambda_0 \beta M_1, & \beta = \alpha \end{cases} \quad (A.9)$$

- the magnitude law

$$1 - F_1(m_\bullet) = (e^{-\beta m_\bullet} - e^{-\beta M_1})/(1-e^{-\beta M_1}) = e^{-\beta m_\bullet}(1+o(1)),\ m_\bullet \gg 1. \quad (A10)$$

Consider the main equation

$$\psi(\pi) = \int_0^M f_1(m)\varphi_F(\psi(\pi)|m)dm, \quad 0 \leq \psi(\pi) \leq 1,$$

where $\psi(\pi)$ is given by (A8).

Since $1 = \int_0 f_1(m)dm$ and $\varphi_F(\psi(\pi)|m) = \phi_F(\lambda(m)(\psi(\pi)-1))$, the equation can be represented as follows

$$\int_M f_1(m)dm + (\psi(\pi)-1) = \int_0^M f_1(m)[\phi_F(\lambda(m)(\psi(\pi)-1))-1]dm, \quad (A11)$$

where $\psi(\pi)$ is defined by the equation

$$\pi = \frac{\varphi_F(\psi|m_\bullet) - \varphi_F(0|m_\bullet)}{1-\varphi_F(0|m_\bullet)} = \frac{\phi_F(\lambda(m_\bullet)(\psi-1)) - \phi_F(-\lambda(m_\bullet))}{1-\phi_F(-\lambda(m_\bullet))} \quad (A12)$$

According to Statement 3, the $\phi_F(w)$ function increases on the semi-axis w<0, and therefore the inverse function $\bar{\phi}_F(\cdot)$ is well-defined. By (A12)

$$(\psi(\pi)-1)\lambda(m_\bullet) = \bar{\phi}_F(\pi + (1-\pi)\phi_F(-\lambda(m_\bullet))) =: C(\pi). \quad (A13)$$

Since $\phi_F(-w) = o(1)$ for w>>1 (see Statement 3) and $\lambda(m_\bullet) \to \infty$ as $m_\bullet \to \infty$,

$$C(\pi) \approx \bar{\phi}_F(\pi).$$

Substituting (A13) into (A11), we have

$$\lambda(m_\bullet)\int_M f_1(m)dm + C(\pi) = \lambda(m_\bullet)\int_0^M f_1(m)[\phi_F(\lambda(m)\lambda^{-1}(m_\bullet)C(\pi))-1]dm$$

or

$$\lambda(m_\bullet)\int_M f_1(m)dm + C(\pi)(1-\int_0^M f_1(m)\lambda(m)dm) = R(M), \quad (A14)$$

where

$$R(M) = \lambda(m_\bullet)\int_0^M f_1(m)[\phi_F(\lambda(m)\lambda^{-1}(m_\bullet)C(\pi))-1-\lambda(m)\lambda^{-1}(m_\bullet)C(\pi)]dm \quad (A15)$$

According to Statement 3 : $|\phi_F(w) - 1 - w| \leq cw^2$, $w < 0$, therefore

$$R(M) \leq cC^2(\pi)\int_0^M f_1(m)\lambda^2(m)dm / \lambda(m_\bullet)$$

$$\leq cC^2(\pi)\int_0^{M_1} f_1(m)\lambda(m)\lambda(M)dm / \lambda(m_\bullet) = cC^2(\pi)ne^{-\alpha(m_\bullet - M)}. \tag{A16}$$

In the case $\alpha = \beta$, direct integration yields

$$R(M) \leq cC^2(\pi)\lambda_0 e^{-\alpha(m_\bullet - M)} \tag{A17}$$

The left-hand part (l.p.) in (A14) can be found explicitly:

$$(\text{l.p. (A14)}) = \lambda_0 e^{-\beta M + \alpha m_\bullet}(1 + o(1))$$

$$+ C(\pi)[(1-n) + n(e^{-(\beta-\alpha)M} \cdot 1_{\beta > \alpha} + (1 - M/M_1) \cdot 1_{\alpha=\beta})(1 + o(1))]. \tag{A18}$$

Using the resulting estimates (A14-A18), we can find an approximate solution to equation (A11, A12).

### 4.1a The case $\alpha < \beta, n < 1$, *AM cluster*.

Substituting

$$M = (\alpha/\beta)m_\bullet + \beta^{-1}\ln(\lambda_0/(1-n))] + x/\beta \tag{A19}$$

in (A17, A18), where $x$ is fixed, we have

$$(1-n)e^{-x} + C(\pi)(1 - n + o(1)) = o(1),$$

that is $C(\pi) = -e^{-x} + o(1)$.

By (A13),

$$\phi_F(-e^{-x} + o(1)) = (\pi + (1-\pi)\phi_F(-\lambda(m_\bullet))).$$

Remembering that $\pi = P(\mu_a \leq M)$, we get

$$P(\mu_a \leq M) = \frac{\phi_F(-e^{-x} + o(1)) - \phi_F(-\lambda(m_\bullet))}{1 - \phi_F(-\lambda(m_\bullet))} \approx \phi_F(-e^{-x}). \tag{A20}$$

In other words

$$\mu_a = (\alpha/\beta)m_\bullet + \beta^{-1}\ln(\lambda_0/(1-n))] + \zeta/\beta,$$

where the random component $\zeta$ has the $\phi_F(-e^{-x})$ distribution on $|x| < \infty$.

From (A20) follows a more accurate approximation

$$P(\mu_a \leq M) \approx \frac{\phi_F(-e^{-\beta M}\lambda(m_\bullet)/(1-n)) - \phi_F(-\lambda(m_\bullet))}{1 - \phi_F(-\lambda(m_\bullet))} 1_{M > \beta \ln 1/(1-n)}. \tag{A21}$$

The limitation M from below here is not surprising, since the approximation (A21) is obtained for the central (peak) values of $\mu_a$.

### 4.1b The case $\alpha = \beta, n \leq 1$, *AM cluster*

In this case $n \approx \beta\lambda_0 M_1 = \beta\lambda_0 K m_\bullet$, and therefore $\lambda_0 = k/m_\bullet = o(1)$.

By setting

$$M = m_\bullet + \beta^{-1} \ln(\lambda_0/(1-n/K))] + x/\beta \qquad (A22)$$

and using (A17), we have

$$(1 - n/K)e^{-x} + C(\pi)(1 - nM/M_1) = O(\lambda_0).$$

Since $M/M_1 = 1/K_1 + O(\ln m_\bullet / m_\bullet) \approx 1/K_1$, we get $C(\pi) \approx -e^{-x}$ again but with a new normalization of $\mu_a$, namely,

$$\mu_a = m_\bullet + \beta^{-1} \ln(\lambda_0/(1-n/K))] + \zeta/\beta, \lambda_0 = k/m_\bullet = o(1),$$

where the random component $\zeta$ has the $\phi_F(-e^{-x})$ distribution.

To obtain the analogue of (A20) we have to replace $n$ in this formula with $\tilde{n} = n/K$.

**4.1c *The case* $\alpha < \beta < 2\alpha, n = 1$, *AM cluster*.**

Let's introduce the notation

$$y = e^{\alpha\mu}, \quad y_0 = e^{\alpha m_\bullet}, \quad M = m_\bullet + x, \quad e^{\alpha x} = \delta. \qquad (A23)$$

Since $\alpha < \beta, n = 1$, we have $\lambda_0 \approx 1 - \alpha/\beta$.

Equation (A14, A15) in the new notation has the form:

$$[\lambda_0(1 + o(1)) + C(\pi)\delta] y_0^{1-\beta/\alpha} \delta^{-\beta/\alpha}$$

$$= \alpha^{-1}\beta\lambda_0 y_0 \int_1^{y_0\delta} y^{-\beta/\alpha} [\phi_F(y/y_0 C(\pi)) - 1 - y/y_0 C(\pi)] dy/y. \qquad (A24)$$

Putting $V = -C(\pi)\delta = -C(\pi)e^{\alpha x}$, $x<0$ and changing the variable under the integral: $y/y_0 \cdot |C(\pi)| = u$, we get

$$\lambda_0 \approx V + (\beta/\alpha - 1)V^{\beta/\alpha} \int_0^V u^{-\beta/\alpha-1} [\phi_F(-u) - 1 + u] du =: \Lambda(V), \qquad (A25)$$

where we replaced $\alpha^{-1}\beta\lambda_0$ with $\beta/\alpha - 1$. Here the lower integral limit was changed from $|C(\pi)|e^{-\alpha m_\bullet}$ to 0. It is acceptable since $\beta/\alpha < 2$ and $\phi_F(-u) - 1 + u = O(u^2), u \to 0$.

Equation (A25): $\Lambda(V) = x$ with respect to $V$ has a single solution $\bar{\Lambda}(x)$ for any $x > 0$. Indeed, the function $f(u) = \phi_F(-u) - 1 + u$ is non-negative (see Statement 3) and therefore $\Lambda(V)$ is an increasing function of $V$. It is clear, that $\Lambda(V)$ increases from 0 to infinity, which guarantees the solvability and uniqueness of equation (A25) for any $\lambda_0 > 0$. Moreover, the non-negativity of the integral term in (A25) entails $\Lambda(V) > V$. Hence $\bar{\Lambda}(\lambda_0) < \lambda_0 \approx 1 - \alpha/\beta < 1$; for small $\lambda_0$ we have $\bar{\Lambda}(\lambda_0) \approx \lambda_0$, because the integral summand in (A25) is of order $V^2$.

Since $\bar{\Lambda}(\lambda_0) = -C(\pi)e^{\alpha x}$, we have by inversion

$$\phi_F(-\bar{\Lambda}(\lambda_0)e^{-\alpha x}) \approx P(\mu_a \leq m_\bullet + x),$$

that is

$$\mu_a = m_\bullet + \alpha^{-1} \ln \bar{\Lambda}(\lambda_0) + \zeta/\alpha, \quad \bar{\Lambda}(\lambda_0) < \lambda_0 \approx 1 - \alpha/\beta < 1,$$

where the random component $\zeta$ has the limit distribution $\phi_F(-e^{-x})$.

As above, we can get a more accurate approximation

$$P(\mu_a \leq M) \approx \frac{\phi_F(-e^{-\alpha M}\lambda(m_\bullet)\bar{\Lambda}(\lambda_0)/\lambda_0) - \phi_F(-\lambda(m_\bullet))}{1 - \phi_F(-\lambda(m_\bullet))} 1_{M > -\alpha^{-1}\ln(\lambda_0/\bar{\Lambda}(\lambda_0))}, \tag{A26}$$

**4.1d The case $0 < 2\alpha \leq \beta, n = 1$, AM cluster**.

Let's set

$$\beta M = 2\alpha m_\bullet - \ln m_\bullet 1_{2\alpha = \beta} + x. \tag{A27}$$

In the case under consideration, $n = 1 \approx \lambda_0 \beta/(\beta - \alpha)$. Therefore the left-hand part of (A14) is

$$(\text{l.p.A14}) = \lambda(m_\bullet)\int_M f_1(m)dm + C(\pi)\int_M f_1(m)\lambda(m)dm$$
$$= \lambda(m_\bullet)e^{-\beta M}(1 + C(\pi)e^{\alpha(M-m_\bullet)}) = \lambda(m_\bullet)e^{-\beta M}(1+o(1)), \tag{A28}$$

since

$$e^{\alpha(M-m_\bullet)} = e^{\alpha x}[e^{-\alpha(1-2\alpha/\beta)m_\bullet}1_{2\alpha < \beta} + (m_\bullet)^{-\alpha/\beta}1_{2\alpha = \beta}] = o(1). \tag{A29}$$

The right-hand part of (A14) is

$$R(M) = \lambda(m_\bullet)\int_0^M f_1(m)[\phi_F(\vartheta(m)) - 1 - \vartheta(m)]dm,$$

where

$$-\vartheta(m) = \lambda(m)\lambda^{-1}(m_\bullet)|C(\pi)| \leq \lambda(M)\lambda^{-1}(m_\bullet)|C(\pi)| = o(1).$$

The o(1) estimate follows from (A29). The continuity of $\phi_F''$ at zero implies that

$$R(M) = \lambda(m_\bullet)\int_0^M f_1(m)\vartheta^2(m)dm \cdot \phi_F''(0)/2 \cdot (1+o(1)). \tag{A30}$$

From equality (A28) and (A30) it follows that

$$\lambda^2(m_\bullet)e^{-\beta M} = C^2(\pi)\int_0^M f_1(m)\lambda^2(m)dm\phi_F''(0)/2 \cdot (1+o(1)).$$

After integration we will have the following:

1) if $2\alpha < \beta$ and $\beta M = 2\alpha m_\bullet + x$ then

$$2(1 - 2\alpha/\beta)/\phi_F''(0) \cdot e^{-x} \approx C^2(\pi),$$

2) if $2\alpha = \beta$ and $\beta M = 2\alpha m_\bullet - \ln m_\bullet + x$ then

$$m_\bullet e^{-x} \approx C^2(\pi)\beta M \phi_F''(0)/2 = C^2(\pi)(2\alpha m_\bullet - \ln m_\bullet + x)\phi_F''(0)/2$$

i.e.,

$$[\alpha\phi_F''(0)]^{-1}e^{-x} \approx C^2(\pi).$$

As above, we can conclude that

$$\beta\mu_a = 2\alpha m_\bullet - \ln m_\bullet 1_{2\alpha=\beta} + \ln A + 2\zeta \ , \ A = 2(\beta\phi_F''(0))^{-1}(\beta - 2\alpha + 1_{\beta=2\alpha}) \ , \quad (A31)$$

where the random component $\zeta$ has the $\phi_F(-e^{-x})$ distribution.

### 4.2 DM($m_\bullet, F$) cluster.

According to Statement1 the DM($m_\bullet, F$) cluster can be considered as an ETAS($\hat{F}$) model, the main characteristics of which are related to the original ones as follows

$$\hat{f}_1(m) = f_1(m)/F_1(m_\bullet)1_{m \le m_\bullet}, \quad (A32)$$

$$\varphi_{\hat{F}}(z|m) = \phi_F(\lambda(m)(zF_1(m_\bullet)-1))/\phi_F(\lambda(m)(F_1(m_\bullet)-1)). \quad (A33)$$

Since $F_1(m_\bullet) \approx 1$ for $m_\bullet \gg 1$, both characteristics converge asymptotically to the characteristics of ETAS($F$). Therefore, it is natural to expect that the limit $\mu_a$-distributions for AM and DM clusters will be the same if the peak value of $\mu_a$ in the AM cluster is much smaller than $m_\bullet$. The method of proof remains the same, though more cumbersome. Therefore, only its main steps are given below.

*The main equation*. Using (A32), (A33) and repeating the previous way, we can find the main equation in a form similar to (A14, A15):

$$\lambda(m_0)\int_M^\infty f_1(m)dm + \hat{C}(\pi)(1-\int_0^M f_1(m)\lambda(m)dm) = \hat{R}(M) \quad (A34)$$

$$\hat{R}(M) = \lambda(m_0)\int_0^{M \wedge m_\bullet} f_1(m)[\frac{\phi_F(\lambda(m)\lambda^{-1}(m_\bullet)\hat{C}(\pi))}{\phi_F(-\lambda(m)(1-F_1(m_\bullet)))} - 1 - \lambda(m)\lambda^{-1}(m_\bullet)\hat{C}(\pi)]dm \ , \quad (A35)$$

where $\hat{C}(\pi) = \lambda(m_\bullet)(\psi(\pi)F_1(m_\bullet)-1)$ and

$$\pi = \frac{\phi_F(\hat{C}(\pi)) - \phi_F(-\lambda(m_\bullet))}{\phi_F(-\lambda(m_\bullet)(1-F_1(m_\bullet))) - \phi_F(-\lambda(m_\bullet))} \ . \quad (A36)$$

### 4.2a,b The cases: $(\alpha < \beta, n < 1)$ and $(\alpha = \beta, n \le 1)$; *DM cluster*.

By assumption, under the $(\alpha, \beta, n)$ conditions, $M$ does not depend on the cluster type. Therefore, the asymptotics of the left-hand sides of equations (A14) and (A34) coincide. For independence of the results from the cluster type it is necessary to check the smallness of the right part of (A35).

We have

$$\beta M = \alpha m_\bullet - \ln m_\bullet 1_{\alpha=\beta} + c \ , \quad (A37)$$

where $c$ is specified in (A19) and (A22). Therefore

$$-\vartheta(m) := \lambda(m)\lambda^{-1}(m_\bullet)|\hat{C}(\pi)| \leq C\lambda(M)/\lambda(m_\bullet)$$

$$= C_1(e^{-\alpha(1-\alpha/\beta)m_\bullet}1_{\alpha<\beta} + m_\bullet^{-1}1_{\alpha=\beta}) = o(1) \tag{A38}$$

$$w(m) := \lambda(m)(1-F_1(m_\bullet)) = -C\vartheta(m)[\lambda_0 e^{-(\beta-\alpha)m_\bullet}1_{\alpha<\beta} + k/m_\bullet 1_{\alpha=\beta}] = o(\vartheta(m)) \tag{A39}$$

where $k = n/(\beta K_1)$. Here we used the relation $\lambda_0 = k/m_\bullet$ at $\alpha = \beta$.

By Statement 3, $|\phi_F(x) - 1 - x| \leq cx^2$, $x < 0$. Hence, the integrand term in square brackets in (A35) admits the estimate

$$|[...]| = |\phi(\vartheta(m)/\phi(-w(m))) - 1 - \vartheta(m)| \leq C_2\vartheta^2(m) .$$

As a result

$$\hat{R}(M) = \lambda(m_\bullet)\int_0^{M\wedge m_\bullet} f_1(m)\vartheta^2(m)dm \leq c\int_0^{M\wedge m_\bullet} f_1(m)\lambda(m)dm \cdot \lambda(M)/\lambda(m_\bullet) = o(1).$$

Consequently, the limit distributions of $\mu_a$ for AM and DM clusters coincide. In particular, in the subcritical regime

$$\hat{C}(\pi) \approx -e^{-x} = -e^{-\beta M}\lambda(m_\bullet)/(1-n).$$

Substitute this relation in (A36). Then, for $(\alpha < \beta, n < 1)$, we will have more accurate approximation

$$P(\mu_a < M) \approx \frac{\phi_F(-e^{-\beta M}\lambda(m_\bullet)/(1-n)) - \phi_F(\lambda(m_\bullet))}{\phi_F(-\lambda(m_\bullet)e^{-\beta m_\bullet}) - \phi_F(-\lambda(m_\bullet))} . \tag{A40}$$

The supports of the measures in this approximate relation do not coincide: $(0, m_\bullet)$ on the left and $(\ln 1/(1-n), m_\bullet + \ln 1/(1-n))$ on the right. This difference practically disappears for $m_\bullet \gg 1$

### 4.2c The case $\alpha < \beta < 2\alpha, n = 1$, *DM cluster*.

Assume that $M = m_\bullet + x$, $x < 0$. Since $n = 1$, $\lambda_0 \approx 1 - \alpha/\beta$ and

$$\text{(l.p.A34)} = \lambda(m_\bullet)e^{-\beta M} + \hat{C}(\pi)\int_M^\infty f_1(m)\lambda(m)dm$$

$$= \lambda(m_\bullet)e^{-\beta M}(1 + \hat{C}(\pi)e^{\alpha x}/\lambda_0) \tag{A41}$$

Using the notation (A38, A39), we represent (A35) as follows:

$$\hat{R}(M) = \hat{R}_1(M) + \hat{R}_2(M) + \hat{R}_3(M) ,$$

where

$$\hat{R}_1(M) = \lambda(m_\bullet)\int_0^M f_1(m)[\phi_F(\vartheta(m)) - 1 - \vartheta(m)]dm ,$$

$$\hat{R}_2(M) = \lambda(m_\bullet)\int_0^M f_1(m)(1/\phi_F(-w(m)) - 1)dm,$$

$$\hat{R}_3(M) = \lambda(m_\bullet)\int_0^M f_1(m)(\phi_F(\vartheta(m)) - 1)(1/\phi(-w(m)) - 1)dm .$$

By (A24, A25)),

$$\hat{R}_1(M) = e^{-\beta M + \alpha m_\bullet} (\beta/\alpha - 1)V^{\beta/\alpha} \int_0^V u^{-\beta/\alpha - 1}[\phi_F(-u) - 1 + u]du(1 + o(1)), \quad (A42)$$

where $V = -\hat{C}(\pi)e^{\alpha x}$. By (A38, A39) and Proposition 1,

$$\hat{R}_2(M) = \lambda(m_\bullet)\int_0^M f_1(m)\lambda(m)(1 - F_1(m_\bullet))dm(1 + o(1)) \approx \lambda_0 n e^{-(\beta-\alpha)m_\bullet}|_{n=1} \quad (A43)$$

and

$$\hat{R}_3(M) = \lambda(m_\bullet)\int_0^M f_1(m)(\phi_F(\vartheta(m)) - 1)(1/\phi(w(m)) - 1)dm$$

$$\leq K\lambda(m_\bullet)\int_0^M f_1(m)|\vartheta(m)|w(m)dm$$

$$= K|\hat{C}(\pi)|\int_0^{m_\bullet + x} f_1(m)\lambda^2(m)dm e^{-\beta m_\bullet} = K_1 e^{-2(\beta-\alpha)m_\bullet} = e^{-(\beta-\alpha)m_\bullet} \cdot o(1). \quad (A44)$$

Finally, combining (A41) with (A42-A44), we get

$$\lambda_0 - V \approx (\beta/\alpha - 1)V^{\beta/\alpha}\int_0^V u^{-\beta/\alpha - 1}[\phi_F(-u) - 1 + u]du + \lambda_0 e^{\beta x}. \quad (A45)$$

In the notation (A25), we have: $\lambda_0 \approx \Lambda(V) + \lambda_0 e^{\beta x}$.

Since $V = -\hat{C}(\pi)e^{\alpha x}$ and the function $\Lambda(V), V \geq 0$ (see (A25)) has inverse $\bar{\Lambda}(\cdot)$, we have $-\hat{C}(\pi) = \bar{\Lambda}(\lambda_0(1 - e^{\beta x}))e^{-\alpha x}$, or

$$P(\mu_a < m_\bullet + x) \approx \phi_F(-\bar{\Lambda}(\lambda_0(1 - e^{\beta x}))e^{-\alpha x}), x \leq 0, \quad (A46)$$

where $x_+ = x \cdot 1_{x \geq 0}$. For small $y$, $\bar{\Lambda}(y) \approx y$ and therefore for small $\lambda_0 = 1 - \alpha/\beta$

$$P(\mu_a < m_\bullet - x) \approx \phi_F(-\lambda(x)F_1(x)).$$

The proof is complete.

### 4.2d The case $2\alpha \leq \beta, n = 1$, DM-cluster.

This case both by proof and by result does not differ from the analogous case for the AM cluster To confirm this, it is enough to compare (A34, A35) with (A28, A30) given

$$\beta M = 2\alpha m_\bullet - \ln m_\bullet 1_{2\alpha = \beta} + x. \quad (A47)$$

Since $n = 1$ and $e^{-\beta(m_\bullet - M)}1_{2\alpha = \beta} = e^{-x}/m_\bullet$,

$$(l.p.(A34)) = \lambda(m_\bullet)e^{-\beta M} + \hat{C}(\pi)e^{-(\beta-\alpha)M} = \lambda(m_\bullet)e^{-\beta M}(1 + o(1)). \quad (A48)$$

So, the asymptotics of (A34) and (A28) are identical. Let's consider (A35):

$$\hat{R}(M) = \lambda(m_\bullet)\int_0^{M \wedge m_\bullet} f_1(m)[\frac{\phi_F(\lambda(m)\lambda^{-1}(m_\bullet)\hat{C}(\pi))}{\phi_F(-\lambda(m)(1 - F_1(m_\bullet)))} - 1 - \lambda(m)\lambda^{-1}(m_\bullet)\hat{C}(\pi)]dm. \quad (A49)$$

Under (A47) condition,

$$-\vartheta(m) = \lambda(m)\lambda^{-1}(m_\bullet)|\hat{C}(\pi)| \leq C\lambda(M)/\lambda(m_\bullet) = o(1),$$

and

$$\lambda(m)(1 - F_1(m_\bullet)) = o(\vartheta^2(m))1_{2\alpha<\beta} + o(\vartheta(m))1_{2\alpha=\beta}.$$

From here

$$\hat{R}(M) = \lambda(m_\bullet) \int_0^M f_1(m)[\vartheta^2(m)\phi_F''(0)/2 + \lambda(m)(1 - F_1(m_\bullet)1_{2\alpha=\beta}]dm(1 + o(1)). \qquad (A50)$$

In the case $2\alpha < \beta$, this relation is identical (A30) and therefore the asymptotic result for both types of clusters is the same.

Assuming $2\alpha = \beta$, we have

$$\hat{R}(M) = \lambda(m_\bullet) \int_0^M f_1(m)[\vartheta^2(m) \cdot \phi_F''(0)/2 + \lambda(m)(1 - F_1(m_\bullet)1_{2\alpha=\beta}]dm(1 + o(1))$$

$$= \lambda(m_\bullet)[\beta M C^2(\pi) \cdot \phi_F''(0)/2 + n]e^{-\beta m_\bullet}(1 + o(1)). \qquad (A51)$$

By equating (A48) and (A51) we obtain

$$[\beta M C^2(\pi) \cdot \phi_F''(0)/2 + n](1 + o(1)) = e^{-\beta(M-m_\bullet)} = m_\bullet e^{-x}.$$

Dividing by $m_\bullet$ and going to the limit, we obtain the same relation as for AM cluster:

$$\alpha C^2(\pi) \cdot \phi_F''(0) = e^{-x}.$$

## Appendix 5.

### Average Cluster Depth

Let $(m_1,...,m_k)$ be a chain of consecutive events in an ETAS cluster and $p_k(m) = P(\nu(m) = k)$. Then the probability density of such chain is

$$f_1(m_1)(1 - p_0(m_1)) \times ... \times f_1(m_{k-1})(1 - p_0(m_{k-1}) \times f_1(m_k)p_0(m_k).$$

Integrating by magnitudes, we obtain the distribution of the chain depth:

$$P(d=k) = (1 - \bar{p}_0)^k \bar{p}_0, \quad k = 0,1,..., \quad \bar{p}_0 = \int_0 p_0(m)f_1(m)dm.$$

If the cluster is initiated by $m_0$-event and contains at least two events, the conditional Average Cluster Depth

$$Ed = 1 + \sum k(1 - \bar{p}_0)^k \bar{p}_0 = 1/\bar{p}_0 = [\int_0 p_0(m)f_1(m)dm]^{-1}.$$

Note, that if $<d>$ is the empirical average of $d$ over all cluster chains, then $E<d> = Ed$ since $d$-expectation is independent of the chain.

The obvious relation

$$1 - p_0(m) \leq \sum_1 k p_k(m) = \lambda(m)$$

entails $\quad Ed \leq (1-n)^{-1}.$ \qquad (A52)

Using the Jensen inequality, we can refine the estimate (A52) for the traditional ETAS model

$$Ed = [\int_0 e^{-\lambda(m)} f_1(m)dm]^{-1} \leq \exp[\int_0 \lambda(m) f_1(m)dm] = e^n, \tag{A53}$$

where $(f_1(m), \lambda(m))$ characteristics can be arbitrary.

This upper bound is valid for any $F = NB(\tau)$ model, because

$p_0(m) = (1 + \lambda(m)/\tau)^{-\tau} > e^{-\lambda(m)}$.